\documentclass[12pt,preprint]{aastex}

\def\ltsima{$\; \buildrel < \over \sim \;$}
\def\gtsima{$\; \buildrel > \over \sim \;$}
\def\lsim{\lower.5ex\hbox{\ltsima}}
\def\gsim{\lower.5ex\hbox{\gtsima}}
\def\lapp{\ifmmode\stackrel{<}{_{\sim}}\else$\stackrel{<}{_{\sim}}$\fi}
\def\gapp{\ifmmode\stackrel{>}{_{\sim}}\else$\stackrel{<}{_{\sim}}$\fi}

\newdimen\minuswidth    

\setbox0=\hbox{$-$}

\minuswidth=\wd0

\catcode`@=\active

\def@{\kern\minuswidth}

\setbox0=\hbox{\rm0}

\shorttitle{Optical emission from J0737-3039}

\shortauthors{Ferraro et al.}

\begin{document} 

\title{Constraining the optical emission from the double pulsar system J0737-3039}

\author{
F. R. Ferraro\altaffilmark{1},
R. P. Mignani\altaffilmark{2,3},
C. Pallanca\altaffilmark{1},
E. Dalessandro\altaffilmark{1},
B. Lanzoni\altaffilmark{1},
A. Pellizzoni\altaffilmark{4},
A. Possenti\altaffilmark{4},
M. Burgay\altaffilmark{4},
F. Camilo\altaffilmark{5},
N. D'Amico\altaffilmark{4,6},
A. G. Lyne\altaffilmark{7},
M. Kramer\altaffilmark{7,8},
R. N. Manchester\altaffilmark{9}
}

\affil{\altaffilmark{1} Dipartimento di Astronomia, Universit\`a degli
  Studi di Bologna, via Ranzani 1, I--40127 Bologna, Italy}
\affil{\altaffilmark{2} Mullard Space Science Laboratory, University
  College London, Holmbury St. Mary, Dorking, Surrey, RH5 6NT, UK}
\affil{\altaffilmark{3} Kepler Institute of Astronomy, University of
  Zielona G\'ora, Lubuska 2, 65-265, Zielona G\'ora, Poland}
\affil{\altaffilmark{4} INAF-Osservatorio Astronomico di Cagliari,
  localit\`a Poggio dei Pini, Strada 54, I-09012 Capoterra, Italy}
\affil{\altaffilmark{5} Columbia Astrophysics Laboratory, Columbia
  University, New York, NY 10027, USA} \affil{\altaffilmark{6}
  Dipartimento di Fisica, Universit\`a di Cagliari, Cittadella
  Universitaria, I-09042 Monserrato, Italy} \affil{\altaffilmark{7}
  Jodrell Bank Centre for Astrophysics, School of Physics and
  Astronomy, The University of Manchester, Manchester M13 9PL, UK}
\affil{\altaffilmark{8} MPI fuer Radioastronomie, Auf dem Huegel 69,
  53121 Bonn, Germany} \affil{\altaffilmark{9} CSIRO Astronomy and
  Space Science, Australia Telescope National Facility, PO Box 76,
  Epping NSW 1710, Australia}
   
\date{February 7, 2012}

\begin{abstract}
We present the first optical observations of the unique system
J0737$-$3039 (composed of two pulsars, hereafter PSR-A and PSR-B).
Ultra-deep optical observations, performed with the {\it High
  Resolution Camera} of the {\it Advanced Camera for Surveys} on board
the {\it Hubble Space Telescope} could not detect any optical emission
from the system down to $m_{\rm F435W}=27.0$ and $m_{\rm F606W}=28.3$.
The estimated optical flux limits are used to constrain the
three-component (two thermal and one non-thermal) model recently
proposed to reproduce the {\it XMM-Newton} X-ray spectrum. They
suggest the presence of a break at low energies in the non-thermal
power law component of PSR-A and are compatible with the expected
black-body emission from the PSR-B surface. The corresponding
efficiency of the optical emission from PSR-A's magnetosphere would be
comparable to that of other Myr-old pulsars, thus suggesting that this
parameter may not dramatically evolve over a time-scale of a few Myr.
\end{abstract} 


\section{INTRODUCTION}
\label{intro}
The binary system J0737$-$3039 is the first and (up to now) the only
double-pulsar system ever detected (Burgay et al.\ 2003; Lyne et
al.\ 2004). It is composed of PSR\ J0737$-$3039A (hereafter PSR-A) and
PSR\ J0737$-$3039B (PSR-B), with spin periods of $22$ ms and $2.7$ s,
respectively, orbiting each other with a $2.5$ hr period along a
mildly eccentric ($\sim 0.09$) orbit.

Radio observations are providing the timing data for exploiting this
formidable system to test General Relativity (Kramer et al.\ 2006,
Kramer \& Stairs 2008). Moreover, they led to the discovery of a
wealth of unprecedented phenomena, like a short ($\sim 30$ s) eclipse
of the radio signal from PSR-A (modulated at full or half period of
PSR-B) when it passes at the superior conjunction (McLaughlin et
al.\ 2004) and a strong modulation of the flux density from PSR-B
along the orbit (Lyne et al.\ 2004). The secular variations of these
peculiar effects have been used to put constraints on the geometry of
the system and on its evolution due to General Relativity (Burgay et
al.\ 2005; Breton et al.\ 2008, Perera et al.\ 2010).  On the other
hand, they clearly demonstrate interaction between PSR-B's
magnetosphere and the spin-down luminosity released by PSR-A in the
form of a relativistic wind of particles and Poynting flux In turn,
this opens the possibility of probing the usually inaccessible
magnetosphere and the local environment of a neutron star (e.g. Zhang
\& Loeb 2004; Lyutikov \& Thompson 2005; Breton et al., in
preparation).

Following the original detection with {\it Chandra} (McLaughlin et
al.\ 2004), J0737$-$3039 has been repeatedly re-observed in the X-ray
band (Pellizzoni et al.\ 2004; Campana et al.\ 2004; Kargaltsev et
al.\ 2006; Chatterjee et al.\ 2007; Possenti et al.\ 2008; Pellizzoni
et al.\ 2008, hereafter P08).  As for the timing, these observations
led to the detection of X-ray pulsations from both pulsars: PSR-A
shows a double-peaked profile with a $\sim 70\%$ pulsed fraction along
the entire orbit (Chatterjee et al.\ 2007), whereas PSR-B displays
pulsations with a complex profile for (at least) a quarter of the
orbit around the ascending node (P08). As for the spectroscopy, the
phased-averaged X-ray spectrum can be satisfactorily fit by
two-component models (a power-law plus a black body model, or a
  double black body model; see P08 and Possenti et al. 2008,
  respectively), while the phase-resolved spectrum is most likely
described in terms of magnetospheric emission from PSR-A and thermal
emission from both the polar caps of PSR-A and the bulk of the PSR-B
surface, irradiated by its companion (P08). If confirmed, the latter
model would represent an additional manifestation of the
  interaction between the PSR-A's spin-down flux and the
magnetosphere of PSR-B. Moreover, the study of the cold black-body
component would allow one to shed light on long-standing questions in
the physics and electro-dynamics of neutron stars, like the thermal
inertia time of the surface and the degree of anisotropy (and the
geometry) of the emission of the spin-down luminosity.

Given the unprecedented nature of the J0737$-$3039 system and the
peculiarities shown in both the radio and X-ray bands, deep
observations of the pulsar system in the optical domain are needed to
complete the study of its phenomenology.  In particular, optical
observations of pulsars are crucial for constraining the tails of both
thermal and non-thermal spectra observed in the soft X-ray regime and
to unveil possible deviations (see Mignani 2011 for a recent
review). Indeed, spectral breaks or turn-offs are found to be quite
common in the thermal/non-thermal optical--to--X-ray spectra of
pulsars (e.g. Mignani et al.\ 2010) and are crucial both to study
anisotropies in the temperature distribution on the surface of the
neutron star and to track the different energy and density
distribution of particles in distinct regions of the pulsar
magnetosphere, where the non-thermal optical and X-ray radiation is
produced.

Here, we present the results of the first, deep, optical observations
of the J0737$-$3039 system.  The data reduction procedures are
presented in \S \ref{data}. Results and discussion are given in \S
\ref{discuss}. Summary and conclusions follow in \S \ref{conclu}.

\section{OBSERVATIONS AND DATA ANALYSIS}
\label{data}
\subsection{Description of the observations}
\label{obs}
The photometric data-set used for this work consists of deep,
high-resolution images obtained with the High Resolution Channel (HRC)
of the Advanced Camera for Surveys (ACS) on board the {\emph Hubble
  Space Telescope} ({\emph HST}).  The HRC consists of one
detector with a pixel-scale of $\sim 0\farcs025$ and a field of view
(FOV) of $\sim 29\arcsec \times 26\arcsec$.  The images have been
obtained on 2005 July 20-21 (Prop. ID 10411, P.I.  Ferraro) in {\em
  ACCUM} mode with a standard multi-point dithering pattern and in two
different bands: six images through the F606W filter
($\lambda=5887.53$\AA; $\Delta \lambda=664.7$\AA; roughly
corresponding to the Johnson's $V$ band) with an exposure time $t_{\rm
  exp}=1250$ s each, and eight images through the F435W filter
($\lambda=4310.97$\AA; $\Delta \lambda=309.5$\AA; corresponding to the
Johnson's $B$ band) with an exposure time $t_{\rm exp}=1300$ s for six
of them, and $t_{\rm exp}=1250$ sec for the remaining two.  Three
short exposures of 10 s each were also acquired in the F606W band.  We
retrieved the data from the \emph{HST} science data archive, after
on--the--fly re-processing by the latest version (v.6.1.0) of the ACS
data reduction pipeline ({\sc calacs}), which applies bias and dark
subtraction, overscan and flat-field correction, and photometry
calibration using the most updated calibration files.
  
As shown in Figures \ref{hrc} and \ref{zoom_hrc}, the pulsar system
(with radio coordinates $\alpha =07^{\rm h} 37^{\rm m} 51\fs248$ and
$\delta = -30^\circ 39\arcmin 40\farcs714$; Lyne et al.\ 2004) is
located at only $\sim 4\arcsec$ from a bright star ($B_J=13.9$
according to the GSC2 catalogue; Lasker et al.\ 2008) also detected in
the 2MASS survey (see Table \ref{tab:2mass}).  In order to prevent it
to severely contaminate the pulsar region, we used the HRC coronograph
to mask this object during the long exposure acquisition.

\subsection{Astrometry}
\label{astrom}
Partially due to the unaccounted proper motions of the reference
stars, the absolute accuracy of the {\emph HST} astrometry can be
worse than the nominal value ($\approx 0\farcs3$). Hence, in order to
precisely locate the pulsar system in the ACS/HRC frame we applied the
following approach.

Since the brightest star is obscured by the coronograph in the long
exposures, for the astrometric re-calibration of the HRC frame we used
the three F606W short images. In particular, we used the drizzled
({\it drz}) images produced by the ACS data reduction pipeline, which
are already corrected for the well known geometric distorsion
affecting the HRC (Pavlovsky et al. 2005). By using the {\sc iraf}
tasks {\tt geomap} and {\tt gregister} we aligned the single images
taking into account shifts and rotations and then, by using the {\sc
  iraf} task {\tt imcombine}, we combined them and applied a
cosmic-ray rejection algorithm.

\begin{figure*}
\includegraphics[height=10cm]{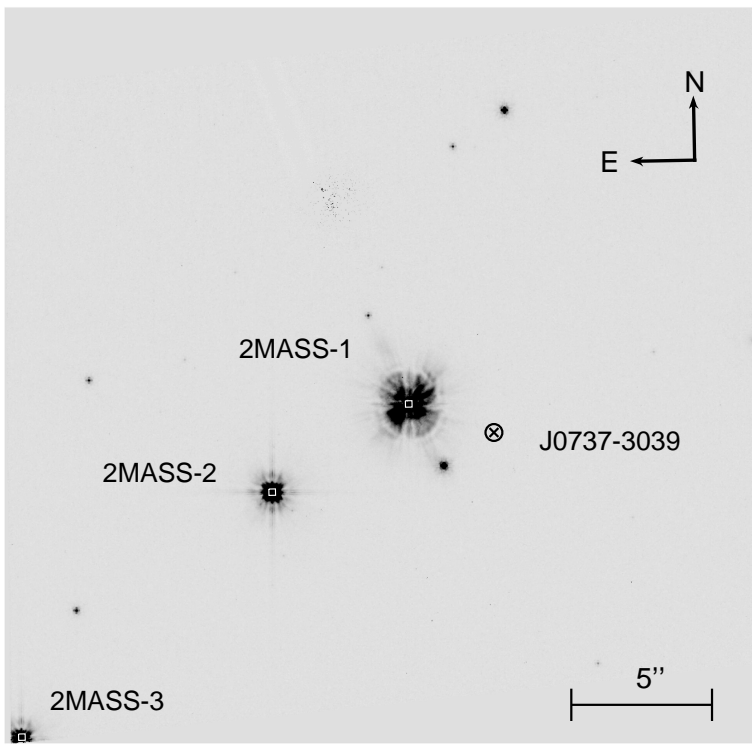}
  \caption{ACS/HRC image of the region around J0737$-$3039 taken
    through the F606W filter (7500 s) after astrometric
    re-calibrations (see \S \ref{astrom}).  The small, white squares
    superimposed to the three brightest stars in the field correspond
    to the 2MASS coordinates given in Table \ref{tab:2mass}. The cross
    corresponds to the estimated location of the J0737$-$3039 system,
    and the circle marks the fiducial uncertainty ($0\farcs3$) on our
    astrometric solution.}
\label{hrc}
\end{figure*}

\begin{figure*}
\includegraphics[height=8.5cm]{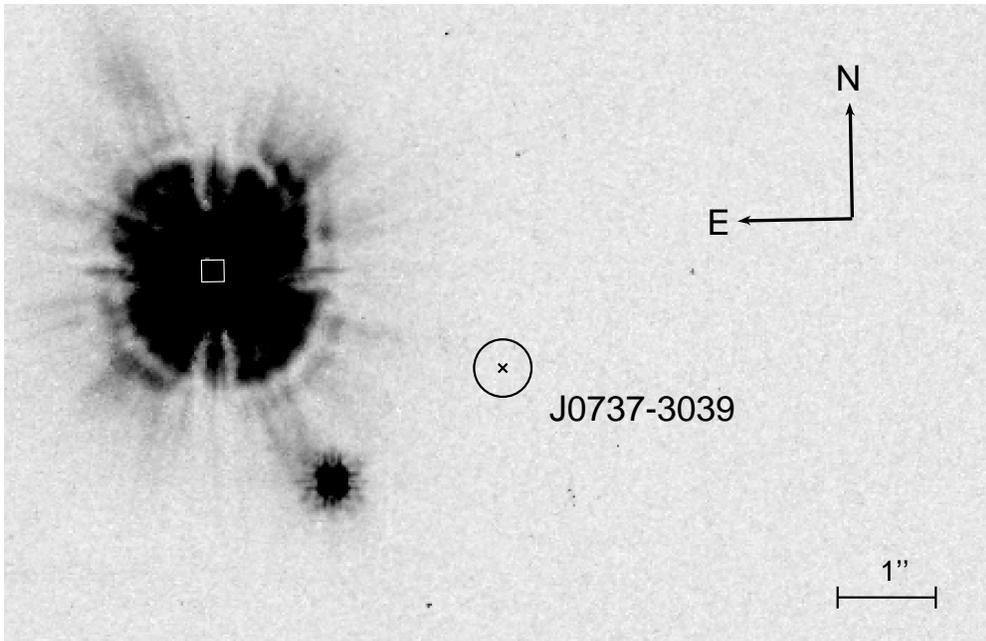}
\caption{Zoomed region around the location of J0737$-$3039 in the
  ACS/HRC image reference frame. The structures observed at the
  position of the bright star 2MASS-1 are the effects of the HRC
  coronograph.}
\label{zoom_hrc}
\end{figure*}

After a quick reduction aimed at just obtaining the instrumental
coordinates of the stars, we cross-correlated our stellar list with
the 2MASS point source catalogue (Skrutskie et al.\ 2006). Only three
2MASS stars have been identified within the FOV of our HRC
observations (see Table \ref{tab:2mass} and Figs. \ref{hrc} and
\ref{zoom_hrc}), their positions (epoch 2000.0) being systematically
offset with respect to the default {\it HST} astrometry by an average
value of $\delta(RA) =-0\farcs21$ and $\delta(Dec) =0\farcs17$, of the
same order of magnitude as found in previous works (e.g., Ferraro et
al.\ 2001).  The image was thus registered on the 2MASS astrometric
grid, which is tied to the international celestial reference frame to
better than 0\farcs15 (Skrutskie et al.\ 2006). Note that the
brightest star (2MASS-1) in the FOV has a measured proper motion:
$\mu_{\alpha\cos\delta}=-42.4 \pm 4.8$ mas yr$^{-1}$ and
$\mu_{\delta}=+30.4 \pm 4.8$ mas yr$^{-1}$ in the UCAC2 catalogue
(Zacharias et al.\ 2004), and $\mu_{\alpha\cos\delta}=-32.5 \pm 4.1$
mas yr$^{-1}$ and $\mu_{\delta}=+29.9 \pm 4.1$ mas yr$^{-1}$ in the
recently released PPMXL catalogue (Roeser et al.\ 2010). This proper
motion would only yield a small displacement (of the order of
$0\farcs15-0\farcs2$) between the epoch of the 2MASS positions
(2000.0) and that of our observations (2005.55).  Taking into account
the rms of the coordinate difference and the uncertainty introduced by
the proper motion of star 2MASS-1, we adopt $0\farcs3$ as a
conservative estimate of the global position error of our astrometry.
The resulting location of the pulsar system in the ACS/HRC frame is
shown in Figs. \ref{hrc} and \ref{zoom_hrc}: it turns out to be
$\approx 3\farcs27$ ($\approx 3\farcs05$ west and $\approx 1\farcs17$
south) away from the bright star 2MASS-1. We note that the pulsar
proper motion, $\mu_{\alpha cos\delta} = -3.82 \pm 0.62$ mas yr$^{-1}$
$\mu_{\delta} = 2.13 \pm 0.23$ mas yr$^{-1}$, only yields a negligible
displacement during the $\sim$ 2 year difference between the epoch of
the reference radio position and that of our \emph{HST} observations.
 
\subsection{Photometry}
\label{phot}
In order to search for even the faintest sources, we performed the
photometric analysis on the median images, by applying to the long
exposures in each filter the same procedure described above. At the
end of the procedure, no object could be detected by eye at the pulsar
position, neither in the F606W nor in the F435W images.

For a more reliable and deeper search of any possible
  counterpart, we estimated the observational detection limit,
  i.e. the minimum counts over the background level that the PSF peak
  of the faintest star should have to be successfully detected by the
  threshold criterium. To this end, we first determined the local
  background level and its standard deviation ($\sigma_{bkg}$) in a
  $1\arcsec\times1\arcsec$ region centered on the nominal pulsar
  system position. Indeed this is a critical task in this particular
  case, since the extended wings of star 2MASS-1 may affect the local
  background level, despite the use of the coronograph.  Following a
  standard approach, the number of counts corresponding to $3\times
  \sigma_{bkg}$ was adopted as the observational detection limit.  We
  then performed a very accurate photometric analysis around the
  nominal position of the pulsar by using {\sc romafot} (Buonanno et
  al.\ 1983), a semi-automatic package that models even complex PSFs
  through an analytic function (Moffat, Gaussian, etc.) combined with
  a numerical residual matrix, and that allows a visual inspection of
  the quality of the fit. In each filter, we first determined the
  shape of the PSF by modeling five bright and non saturated stars in
  the HRC FOV, by using a Moffat function combined with the numerical
  residual matrix. We then used such a PSF to search for sources at
  more than $3\times\sigma_{bkg}$ in a region centered on the pulsar
  system position. No signal was found above this limit even when
  applying a Gaussian smoothing to the images over a $3\times3$ pixel
  cell.  We therefore conclude that J0737$-$3039 is undetected in our
  ACS/HRC images.

The magnitude of the faintest detectable object has been computed as
the magnitude of a star having the adopted PSF and a number of counts
at peak equal to $3\times\sigma_{bkg}$.  By adopting the Charge
Transfer Efficiency correction (Maybhate et al.\ 2010) and the
absolute flux calibration (from the updated {\em PHOTFLAM}
coefficients available in the most recent {\sc synphot} tables
maintained at StScI) our $3 \sigma_{bkg}$ upper limits turned out to
correspond to $m_{\rm F435W}=27.0$ and $m_{\rm F606W}=28.3$ in the
STmag photometric system.\footnote{As a sanity check, we used the
    empirical PSF to simulate two artificial stars of different
    magnitude ($m_{\rm F606W}=28.5$ and 28.3) directly in the images, at
    the nominal position of the pulsar system. By re-performing the
    entire photometric analysis on these images, we found that only
    the brighter star could be successfully detected and measured.}

We finally transformed these magnitude limits into upper limits to the
unabsorbed spectral flux. To do this, we first evaluated the
extinction in the direction of the pulsar system, by using the
relation of Predehl \& Schmitt (1995) between the hydrogen column
density $N_{\rm H}$ and the extinction coefficient $A_V$.  While this
relation is affected by uncertainties for close objects, due to the
problems of modeling the interstellar medium at small distance from
the Sun where microstructures weight more, we checked that, when
using, e.g., the relation of Paresce (1984), the resulting extinction
corrections are consistent within 0.01 magnitudes.  Assuming $N_{\rm
  H} = 6.9^{+1.5}_{-1.1}\times10^{20}$ cm$^{-2}$ (P08)\footnote{The
  quoted uncertainties in $N_{\rm H}$ translates in a 0.08 mag
  uncertainty in $A_V$. } and $A_V=3.12\times E(B-V)$, we obtained a
colour excess $E(B-V)=0.123$. This was used to compute the extinction
coefficients appropriate for the central wavelength of the adopted
filters (Fitzpatrick 1999), thus obtaining $A_{\rm F435W}=0.52$ and
$A_{\rm F606W}=0.35$.  Using the standard relation for the STmag
system (mag$_\lambda= -2.5 \log F_\lambda -21.1$), we obtained the
monochromatic fluxes $F_{\rm F435W}= 9.3\times 10^{-20}$ erg cm$^{-2}$
s$^{-1}$ \AA$^{-1}$ and $F_{\rm F606W}= 2.4\times 10^{-20}$ erg
cm$^{-2}$ s$^{-1}$ \AA$^{-1}$.  The corresponding quantities in units
of the frequency have been obtained by considering the effective
wavelength of the two pass-bands: $\lambda_{\rm F435W}=4310.9702$
\AA\ and $\lambda_{\rm F606W} = 5887.536$ \AA.  Finally, according to
standard unit conversions (e.g. Zombeck 2007), we derived the values
of the unabsorbed spectral fluxes $F_{\rm F435W}=8.7\times10^{-5}$ keV
cm$^{-2}$ s$^{-1}$ keV$^{-1}$, and $F_{\rm F606W}=4.2\times10^{-5}$
keV cm$^{-2}$ s$^{-1}$ keV$^{-1}$.
 
\section{DISCUSSION}
\label{discuss}
The derived upper limits to the optical flux in the F435W and F606W
filters can be used to characterize the optical emission of the
  J0737$-$3039 system (\S 3.3), as well as to constrain the spectral
  model used to fit the X-ray data (\S 3.2), which is also briefly
  reported below (\S 3.1) for the sake of clarity (a detailed
  discussion of the X-ray spectra of the system is given in P08).

\subsection{The X-ray spectrum}
\label{xspec}
By exploiting deep XMM-{\it Newton} observations performed on October
2006, P08 could perform phased-resolved spectroscopy of J0737$-$3039
and found that a model with three components, a power law (PL) and two
black-bodies (BBs), is required to fit both the phase-resolved spectra
and the energy dependence of the pulsed fraction of PSR-A.  According
to this model, the PL (with photon index $\Gamma =
3.3^{+0.1}_{-0.2}$) is responsible for most of the pulsed emission
originating in the magnetosphere of PSR-A (e.g. Zhang \& Cheng 1999;
Zhang \& Harding 2000).  The hotter and fainter BB (HBB, with
$kT=134^{+17}_{-14}$ eV) has a $\sim 100$ m emitting radius
\footnote{The quoted radii of the BB components have been obtained by
  P08 under the assumption of a distance of 0.5 kpc, inferred from the
  dispersion measure of the two pulsars (Lyne et al.\ 2004) and a
  model for the distribution of electrons in the Galaxy (Cordes \&
  Lazio 2002). We used this distance as a reference value. Within the
  uncertainties in spectral modeling and fitting, the resulting
  parameters are compatible with those derived from the still
  preliminary determinations of the annual geometric parallax of the
  system (0.2-1 kpc, Kramer et al 2006; $\sim 1$ kpc, Deller et
  al. 2009).} and it is likely associated with the polar caps of
PSR-A, re-heated by back-flowing particles accelerated in its
magnetosphere (e.g. Cheng \& Ruderman 1980; Arons 1981; Zavlin \&
Pavlov 1998; Zavlin 2006; Bogdanov 2006). The colder and brighter BB
(CBB, with $kT=32^{+5}_{-4}$ eV) is associated with an emitting radius
of $\sim 15$ km and plausibly arises from the bulk of the surface of
one of the two neutron stars.  The most likely explanation is that it
is due to thermal radiation coming from the bulk of PSR-B's surface,
after the conversion of $\sim 2\%$ of the spin-down luminosity of
PSR-A intercepted by PSR-B at suitable orbital phases (see e.g. Zhang
\& Loeb 2004; Lyutikov \& Thompson 2005). This is also consistent with
the observed soft pulsed flux from PSR-B (P08).

\subsection{Optical vs. X-rays}
In order to use the derived optical flux upper limits for constraining
the spectral model discussed above and getting new clues on the
physics of the system, it is necessary to extrapolate the PL and BB
components into the optical domain.  In agreement with the evidence
found in the optical spectra of a few Myr old pulsars (Mignani 2011),
a contribution in the optical band from the magnetospheric emission
can be expected. Since PSR-A's spin-down energy is $\sim 3000$ times
higher than PSR-B's one ($\dot{E}_A=5.9\times 10^{33}$ erg s$^{-1}$;
$\dot{E}_B=1.7\times 10^{30}$ erg s$^{-1}$; Lyne et al. 2004), we
assume that the former would dominate any magnetospheric optical
emission from the system, as it does in the X-rays.  On the other
hand, we note that the actual broad-band spectrum of PSR-A is, in
principle, compatible with comptonised BB models (see e.g. Nishimura
et al.\ 1986), which are not expected to provide a non-thermal
contribution at optical wavelengths. While further X-ray observations
are required to improve the photon count statistics and constrain
complex multi-component phase-resolved spectral models (including
comptonisation scenarios), here we conservatively take into account
the possibility of a contribution of the non-thermal component even at
low energies.

\begin{figure*}
\begin{center}
\includegraphics[height=10cm]{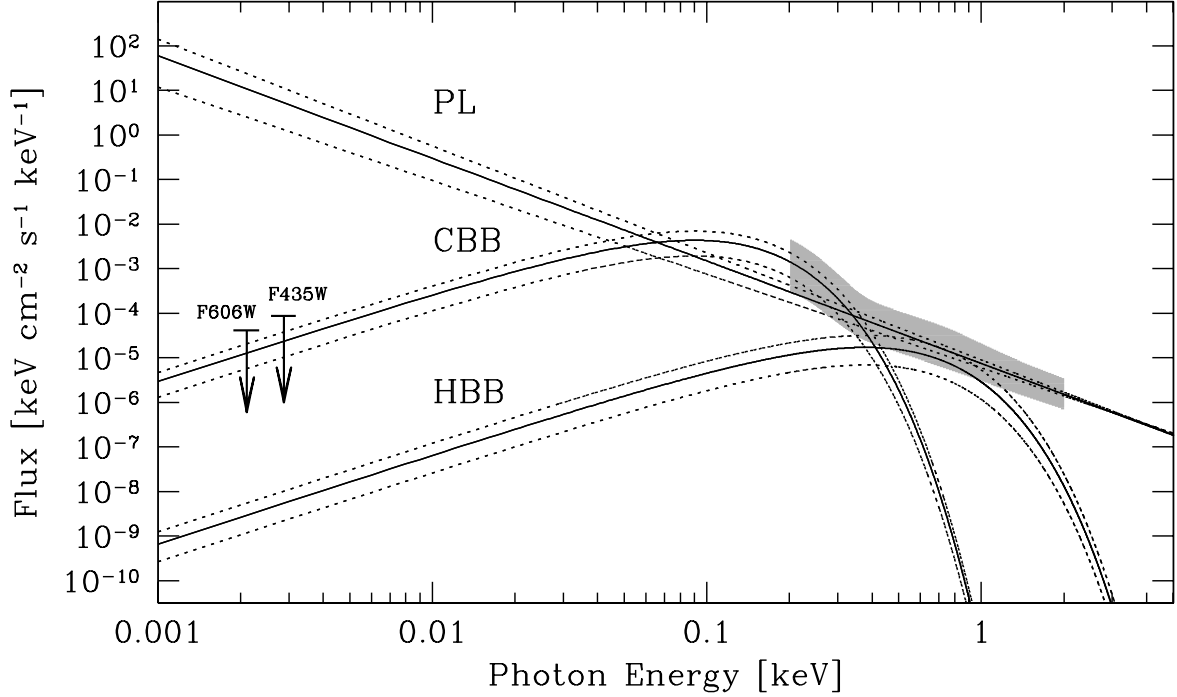}
\end{center}
\caption{Extrapolation into the optical regime of the model that best
  fits the XMM-{\it Newton} data of the pulsar system: a cooler BB and
  a hotter BB plus a PL.  The solid lines are the best fit model
  components, while the dot-dashed lines correspond to the 90\%
  confidence level uncertainties of the best fit.  The grey
    region marks the location and uncertainties of the X-ray
    measurements.  The two arrows indicate the F435W and F606W-band
  upper limits obtained from the {\em HST} data. Since the uncertainty
  on the extinction correction derived from the assumed $N_{\rm H}$
  implies only a difference of $\la 7\%$ in flux, we have neglected it
  when plotting the upper limit values.  }
\label{spec}
\end{figure*}

The extrapolation of the PL and BB components into the optical domain
is shown in Fig. \ref{spec}, together with the optical flux upper
limits computed in \S \ref{phot}.  While the expected contribution of
the HBB is at least four orders of magnitude lower than the derived
optical values, the extrapolation of the CBB component only slightly
undershoots the optical flux upper limits, the difference being
smaller than a factor of 10 after accounting for the 90\% confidence
level uncertainty of the spectral parameters. Finally, the
extrapolated non-thermal PL component of the X-ray spectrum overshoots
the optical flux upper limits by about six orders of magnitude.

As a first consideration, our optical upper limits indicate that
either the PSR-A magnetosphere does not provide a non-thermal
contribution at low energies (as expected in the comptonised BB
model), or a break at these wavelengths is present.\footnote{Note that
  if the X-ray PL component was extrapolated to optical wavelengths
  with the same index, it would imply an emitted power in the optical
  band which is a significant fraction (0.1-0.2) of the measured
  PSR-A's spin-down luminosity, in striking contrast with what
  observed so far in the pulsar population (see \S 3.3).  This
  argument further supports the presence of a spectral break between
  the optical and the X-ray domains.}  Indeed, such breaks are quite
common in the optical--to--X-ray magnetospheric spectra of pulsars
above a wide age range (e.g., Mignani et al.\ 2010) and are probably
indicators of a complex particle density and velocity distribution in
the neutron star magnetosphere. However, in order to be compatible
with the expected optical emission from the CBB component (see
Fig. \ref{spec}), the expected spectral break in PSR-A's spectrum has
to be very sharp if it occurs at energies lower than $\sim 0.1$
  keV.

In the case of a null or negligible optical contribution of the PL
component, the fact that the measured upper limits are consistent with
the expectations from the combined BB components represents a partial
validation of the model and the adopted spectral parameters.

By neglecting the contribution of PSR-A's polar caps (i.e., the HBB
component), the comparison between the optical upper limits and the
expected CBB contribution more quantitatively suggests that the
PSR-A's magnetospheric optical emission cannot exceed the surface
thermal emission from PSR-B by a factor larger than $\sim 5$.
Conversely, by assuming that the PL spectral break is such to make any
magnetospheric optical emission negligible with respect to the thermal
one, we can derive a qualitative upper limit to the temperature of the
bulk of PSR-A surface (i.e. away from the hot polar caps) to account
for the possible flux excess with respect to the expected CBB
contribution.  For instance, in order to be compatible with the
non-detection of a third BB component (cooler than the CBB) in the
observed X-ray spectrum, for an emitting radius of 15 km, the bulk of
PSR-A surface should be at a temperature $kT \la 20$ eV. This is
consistent with its spin-down age ($\sim 200$ Myr) and a passive
cooling scenario.

Spectral data in the ultraviolet would be useful to constrain the
presence and energy of a break at low energy in the PL spectrum of
PSR-A, as well as the temperature of the CBB component produced by
PSR-B. Unfortunately, J0737$-$3039 has not been observed by {\emph
  GALEX} (Martin et al.\ 2005) and the XMM-{\it Newton} Optical/UV
monitor camera (Mason et al.\ 2001) has been used only in the $V$
band, while the relatively recent discovery of the system prevented
its observation within the pulsar survey carried out by {\emph EUVE}
(Korpela \& Bowyer 1998).

\subsection{Optical emission efficiency}
The F606W upper limit corresponds to an extinction-corrected optical
luminosity for the pulsar system $L_{\rm F606W} \la 4.8 \times 10^{26}
d_{500}^2$ erg s$^{-1}$ (where $d_{500}$ is its distance in units of
500 pc), with a $\la 7\%$ uncertainty due to the error on the
extinction correction computed from $N_{\rm H}$ (see \S \ref{phot}).
In the limit case that PSR-A's emission is mostly magnetospheric, and
under the assumption that its contribution to $L_{\rm F606W}$ is at
most a factor of 5 larger than that of the PSR-B's thermal emission
(see above), we can derive the optical emission efficiency:
$\eta_{opt,A}\equiv L_{\rm F606W}/\dot{E}_A \la 8 \times
10^{-8}$~$d_{500}^2$.

Such a value of $\eta_{opt}$ implies that the optical emission
efficiency of PSR-A might still be compatible with that of most
pulsars with strong magnetospheric emission (e.g., Zharikov et
al.\ 2006), accounting for the uncertainties on the distance and for
the difference in the reference spectral band\footnote{The
  efficiencies quoted in Zharikov et al.\ (2006) are computed in the B
  band.}. This might indicate that the optical emission efficiency
does not dramatically evolve with the pulsar's spin-down age and
remain relatively high for pulsars as old as $\sim 200$ Myr, like
PSR-A (Lorimer et al.2007). Unfortunately, given the small
number of pulsars identified in the optical domain, of which only a
handful are older than 1 Myr (see Mignani 2011), it is very difficult
to outline a clear evolutionary pattern.  For instance, for the 3 Myr
old PSR\, B1929+10 (Mignani et al.\ 2002) the optical luminosity, {\em
  if} entirely magnetospheric, would imply $\eta_{opt} \sim 4.7 \times
10^{-7}$, i.e.  above the upper limit derived for PSR-A. Under the
same assumption, for the $\sim 4$ Myr old PSR\, B1133+16 (Zharikov et
al.\ 2008) and for the 17 Myr old PSR\, B0950+08 (Zharikov et
al.\ 2004) we derive $\eta_{opt} \approx 10^{-6}$ and $\sim 1.3 \times
10^{-6}$, respectively, while for the $\sim$ 166 Myr old PSR\,
J0108$-$1431 (Mignani et al.\ 2008; 2011), whose optical
identification is however unconfirmed, $\eta_{opt}$ would be as high
as $\approx 5 \times 10^{-5}$.

Surprisingly enough, these values are comparable to those of much
younger and brighter pulsars, like the Crab and PSR\,
B0540$-$69. However, we warn here that, while the optical emission of
young pulsars is entirely magnetospheric (e.g., Mignani et al.\ 2010),
that of the old pulsars partially (if not entirely) arises from the
neutron star surface.  Thus, the above quoted values of $\eta_{opt}$
are rather uncertain, and optical spectra would be needed to
disentangle the contributions of the magnetospheric and thermal
components.  Indeed, in the case of PSR\, J0108$-$1431, whose
spin-down age is comparable to that of PSR-A, the optical spectrum of
its (candidate) counterpart is compatible with the Rayleigh-Jeans tail
of a $kT\sim 0.43$~$d_{240}^2$ eV BB for a 13 km neutron star radius
(Mignani et al.\ 2008; 2011), implying that the bulk of the optical
emission arises from the neutron star surface and that the actual
emission efficiency is much lower than $\approx 5 \times 10^{-5}$.  On
the other hand, since PSR-A has a $\approx 1000$ times larger
spin-down energy, its magnetospheric emission is probably much
stronger than that of PSR\, J0108$-$1431 and, possibly, it is dominant
over the thermal component.  Thus, the derived luminosity upper limits
do not rule out that the (yet undetected) optical emission from the
pulsar system could be of magnetospheric origin and almost entirely
associated with PSR-A.

\section{Summary and conclusions}
\label{conclu}
We performed the first ever follow-up optical observations of the
J0737$-$3039 system. Our deep ACS/HRC observations could not detect
optical emission from the system down to limiting magnitudes of
$m_{\rm F435W}=27.0$ and $m_{\rm F606W}=28.3$, corresponding to upper
limits to the unabsorbed spectral fluxes $F_{\rm
    F435W}=8.7\times10^{-5}$ keV cm$^{-2}$ s$^{-1}$ keV$^{-1}$ and
  $F_{\rm F606W}=4.2\times10^{-5}$ keV cm$^{-2}$ s$^{-1}$ keV$^{-1}$.
  In the framework of the PL+2BB model that best fits the XMM-{\it
  Newton} spectrum (P08), the derived flux upper limits suggest a
break at low energies in the PL component, similar to what observed in
other pulsars (e.g., Mignani et al.\ 2010), and are consistent with
the BB temperature and emitting area of PSR-B.  The upper limit to the
optical luminosity of the pulsar system ($L_{\rm F606W} \la 4.8 \times
10^{26} d_{500}^2$ erg s$^{-1}$) implies that the optical emission
efficiency of PSR-A, which has the highest spin-down power, would be
comparable to that of Myr-old pulsars, suggesting that their optical
magnetospheric emission does not dramatically evolve over a time-scale
of a few Myr.  Deep observations in the ultraviolet with the {\emph
  HST} and its refurbished instrument suite will probably offer higher
chances of detection and, thus, of better constraining the thermal
spectrum of PSR-B and the magnetospheric spectrum of PSR-A.

\acknowledgements We warmly thank the referee for the critical
  reading of the manuscript and the useful comments. This research is
part of the project {\it COSMIC-LAB} funded by the European Research
Council (under contract ERC-2010-AdG-267675).  The financial
contribution of the Agenzia Spaziale Italiana (under contract ASI-INAF
I/009/10/0) and of the Italian Istituto Nazionale di Astrofisica
(INAF; under contract PRIN-INAF 2008) is also acknowledged.

\newpage
\begin{table}
\caption{Absolute positions of the three 2MASS stars (Skrutskie et
  al.\ 2006) used to compute the astrometric solution in the ACS/HRC
  frame (\S 2.2).}
\begin{center}
\begin{tabular}{c | c | c }
\hline
NAME & R.A. & Dec.  \\
     &   (J2000) & (J2000)  \\
\hline
2MASS-1 & 07 37 51.48 & -30 39 39.57\\
2MASS-2 & 07 37 51.85 & -30 39 43.41 \\
2MASS-3 & 07 37 52.51 & -30 39 54.05\\
\hline
\end{tabular}
\label{tab:2mass}
\end{center}
\end{table}

\newpage

\newpage

\newpage

\end{document}